\documentclass[useAMS,usenatbib,referee]{biom}
\usepackage{amsfonts}
\usepackage{graphicx}

\def\*#1{\bm{#1}}
\def\bSig\mathbf{\Sigma}

\title[Non-asymptotic inference for multivariate change point detection]{Non-asymptotic inference for multivariate change point detection}

\author{Ian Barnett\emailx{ibarnett@pennmedicine.upenn.edu} \\
Department of Biostatistics, University of Pennsylvania, Philadelphia, PA, U.S.A.}

\begin{document}


\date{{\it Received May} 2018. {\it Revised ???} ????.  {\it
Accepted ???} ????.}



\pagerange{\pageref{firstpage}--\pageref{lastpage}} 
\volume{??}
\pubyear{????}
\artmonth{???}


\doi{10.1111/j.1541-0420.2005.00454.x}


\label{firstpage}


\begin{abstract}
Traditional methods for inference in change point detection often rely on a large number of observed data points and can be inaccurate in non-asymptotic settings. With the rise of mobile health and digital phenotyping studies, where patients are monitored through the use of smartphones or other digital devices, change point detection is needed in non-asymptotic settings where it may be important to identify behavioral changes that occur just days before an adverse event such as relapse or suicide. Furthermore, analytical and computationally efficient means of inference are necessary for the monitoring and online analysis of large-scale digital phenotyping cohorts. We extend the result for asymptotic tail probabilities of the likelihood ratio test to the multivariate change point detection setting, and demonstrate through simulation its inaccuracy when the number of observed data points is not large. We propose a non-asymptotic approach for inference on the likelihood ratio test, and compare the efficiency of this estimated p-value to the popular empirical p-value obtained through simulation of the null distribution. The accuracy and power of this approach relative to competing methods is demonstrated through simulation and through the detection of a change point in the behavior of a patient with schizophrenia in the week prior to relapse.
\end{abstract}

%



\maketitle


%

\section{Introduction}
\label{sec:intro}

The primary aim of this work is to accurately estimate the tail probabilities of multivariate change point detection likelihood ratio test statistics without reliance on asymptotics or computationally intensive simulation under the null distribution. The context where this problem emerges is in the growing number of studies that rely on \textit{digital phenotyping}, defined by \citet{torous2016new} to be the ``moment-by-moment quantification of the individual-level human phenotype in-situ using data from smartphones and other personal digital devices." Digital phenotyping is being increasingly used to study human behavior, primarily for the purposes of monitoring vulnerable populations with psychiatric disorders \citep{onnela2016harnessing,torous2017new,insel2017digital}. A typical digital phenotyping study will have patients install an application on their own personal smartphone that will collect sensor data in the background during everyday use. This data can be used to calculate behavioral features related to patient location, mobility, activity, sociability, and sleep with high granularity over long periods of time. Considering these behavioral features for a single patient as a multivariate time series, change points in this time series may correspond with an alteration in the patient's behavior or an escalation of symptoms leading up to an adverse event such as relapse or suicide. Because change point detection is necessary with only a few observed data points post-change point, asymptotic methods of inference are not applicable

While there is a rich body of research devoted to the development of powerful methods for multivariate change point detection, many treatments focus only on identifying the most likely location of change points while failing to address the corresponding problem of statistical inference without the use of Monte Carlo simulation of the null distribution. For example, cumulative sum control chart (CUSUM) based methods \citep{cho2016change,cho2015multiple}, a pairwise likelihood approach \citep{ma2016pairwise}, generalizations of likelihood-based cost functions \citep{killick2012optimal}, methods that allow for change points in the covariance of features \citep{lavielle2006detection,kirch2015detection}, nonparametric methods \citep{preuss2015detection} such as a method based on the sample divergence \citep{matteson2014nonparametric}, a dimension-reducing method that allows for missing data \citep{xie2013change}, and a method capable of relaxing the assumption that the change point occurs simultaneously across the multivariate time series have been developed \citep{bardwell2018most}. While these methods are designed to be powerful to detect change points under certain alternatives, in each case the problem of non-asymptotic statistical inference is all but ignored, often relying exclusively on Monte Carlo simulation of the null distribution for hypothesis testing. While simulation of the null distribution may be acceptable in the application settings they consider, in this paper we quantify how such inference can be prohibitively computationally expensive in cases where the significance level is small. The problem of small significance levels is present in smartphone-based digital phenotyping studies because of the corrections for the multiple comparisons necessary due to repeated online analyses over long periods of follow-up. To address this issue, in this paper we develop a method of non-asymptotic inference in one of the most popular and fundamental methods for multivariate change point detection in the mean: the likelihood-ratio test.

Estimation of the asymptotic tail probabilities and p-values for likelihood-ratio change point detection test statistics has been studied, with a comprehensive asymptotic treatment of a variety of change point problems demonstrated in \citet{siegmund2013sequential}. With sufficient follow-up time digital phenotypes can be approximated by Brownian motion due to the weak convergence of stochastic processes, from which tail probabilities can be obtained \citep{billingsley2008probability}.  For univariate processes, asymptotic approximations of these tail probabilities have been developed for detecting one change point \citep{woodroofe1982nonlinear} as well as two change points, i.e. a change interval \citep{siegmund1992tail,hogan1986large}. The extension to multivariate processes was explored in \citet{james1987tests} for single change points and in \citet{zhang2010detecting, siegmund2011detecting} for change point intervals, where change point detection was used to detect copy number variation and the use of asymptotics was appropriate. In digital phenotyping studies, many fewer observations are available. For example, if behavioral features are gathered every day, then using change point detection to prompt an intervention will only be useful if the change point is detected within a few days after it occurs. The problem of change point detection in non-asymptotic settings is considered by the CUSUM procedure used for statistical process control \citep{page1954continuous}, but non-asymptotic treatments of likelihood-ratio tests for multivariate change point detection problems are relatively unexplored. An exact tail probability for the multivariate case has been developed using infinite series of Bessel functions in \citet{kiefer1959k}, but this approach searches the entire process for a change point, whereas we are only interested in focused testing for a change point in a patient's relatively recent history.

The rest of the paper is organized as follows. We first present the mean change point model in Section \ref{sec:model}. In Section \ref{sec:methodsp} we derive the asymptotic probabilities of the multivariate change point detection and demonstrate its poor accuracy in non-asymptotic setting as compared to our proposed method for non-asymptotic inference. In Section \ref{sec:pvalacc} we compare the efficiency of our proposed p-value estimate to that obtained from Monte Carlo simulation of the null distribution, both theoretically and through simulation. Proofs are left to the appendix. In Section \ref{sec:power} power simulations are used to compare competing methods, and in Section \ref{sec:realdata} we demonstrate the potential of multivariate change point detection in digital phenotyping studies by applying our method to detect a change point in the digital phenotypes of a patient with schizophrenia in the week prior to relapse.


\section{Model}
\label{sec:model}

We consider observed data $Y_{ij}$ that are distributed according to:
\begin{equation}\label{model}
Y_{ij} = \mu_{i} + I_{\{j > k\}}\delta_i  + \epsilon_{ij}
\end{equation}
for features $i =1,\ldots,q$ and observations $j = 1,\ldots,n$ ordered sequentially in time, where $I_{\{x\}}$ is the indicator function for $x$,  and $\epsilon_{ij}$ are independent and identically distributed standard normal errors. A similar change point detection model is considered in \citet{zhang2010detecting} where change intervals $(k_0,k_1]$ of some minimum width such that $1 \leq k_0 < k_1 \leq n$ specify when there is a shift of $\delta_i$ in the mean of feature $i$. Model \ref{model} represents the special case of this when $k_1=n$.  The reason for this restriction on $k_1$ is because when using change point detection to prompt interventions in vulnerable populations, it is only important to identify recent behavioral changes. Effective interventions require a prompt response to the behavioral change, and intervals of changed behavior in a patient's past ($k_1<n$) imply that the patient's behavior eventually returned back to its normal state, making an intervention no longer necessary.

For feature $i$, testing for a change point at time $k$ corresponds with $H_0: \delta_i=0$ and the two-sided alternative $H_1: \delta_i\neq0$. Letting $\bar{Y}_{i} = \sum_{j=1}^nY_{ij}/n$, the likelihood-ratio based test statistic proposed by \citet{olshen2004circular} and \citet{zhang2010detecting} for a change point in a single feature $i$ at time $k$ is:
\begin{equation}
U_{ik} = \frac{(\sum_{j=k+1}^{n}Y_{ij})-\bar{Y}_{i}(n-k)}{\sqrt{k(n-k)/n}}
\end{equation}

The nature of this test statistic is intuitive: if a shift in the mean of $Y_{ij}$ occurs at time $k$, then the sum across the observed data after the change point should be substantially different compared to the observed average over all of the data, $\bar{Y}_{i}$. To test for a simultaneous change point across all $p$ features at time $k$, we consider the test statistic: 

\begin{equation}\label{Zk}
Z_k = \sum_{i=1}^q U_{ik}^2
\end{equation}
For each $k$, $Z_k \sim \chi_q^2$ under $H_0$, which corresponds to the case when $\delta_1=\cdots=\delta_q=0$. As the change point location is unknown, we search for the most likely change point location in the interval $[m_0,m_1]$ by considering the test statistic:

$$Q = \max_{n-m_1\leq k\leq n-m_0}Z_k$$
We reject $H_0$ for large values of $Q$. 

\section{Methods for estimating p-values}
\label{sec:methodsp}

For large $n$, \citet{siegmund2013sequential} derives an approximation for the p-value $\mbox{pr}(Q\geq b^2)$ in the univariate case ($q=1$), which can be extended to arbitrary $q$ features as follows:

\begin{theorem}
Suppose that $b\rightarrow \infty$, $m_1 = o(n)$, $m_0\rightarrow \infty$, and $m_1\rightarrow \infty$ such that $bm_1^{-1/2}=c_1<c_0=bm_0^{-1/2}$ for constants $c_0$, $c_1$. Then under $H_0$
\begin{equation}\label{pvalasymp}
\mbox{pr}(Q\geq b^2)\approx 2^{-q/2}\left[\Gamma\left(q/2\right)\right]^{-1}\ln\left(m_1/m_0\right)b^q\exp\left(-b^2/2\right)
\end{equation}
\end{theorem}

For the purposes of smartphone monitoring with daily observations, we are frequently interested in the case where, say, $m_0=0$ and $m_1 = 6$. This corresponds to looking for a change point in patient behavior over the previous week. Because $m_0$ and $m_1$ are fixed to be small, even if a person is followed for a long time ($n\rightarrow \infty$), the asymptotic approximation will not improve. Inference must instead rely on a non-asymptotic approach.

As an alternative to this asymptotic approximation, we detail how to estimate  $\mbox{pr}(Q\geq b^2)$ without assuming large $b$, $n$, $m_0$, and $m_1$ and without requiring computationally expensive simulation of the null distribution. A useful approximation is
\begin{equation}\label{1storder}
\mbox{cor}(Z_{j_1},Z_{j_2}) \approx (n-j_2)/(n-j_1) \;\;\; (1 \leq j_1 < j_2 \leq n).
\end{equation}
When we condition on $\bar{Y}_{i}$ this correlation is exact, implying that the approximation holds well for $n-m_1 \leq j_1 < j_2 \leq n-m_0$ when $m_1$ and $m_0$ are close to $0$ as is the case for change points locations of interest in digital phenotyping studies. Letting $F_q(\cdot)$ be the cumulative distribution function of a $\chi_q^2$ random variable, we define $Z_k^{*}=\Phi^{-1}\{F_q(Z_k)\}$ to be a transformation of $Z_k$ that is normally distributed under $H_0$. Letting
$$Q^{*} = \max_{n-m_1\leq k\leq n-m_0}Z_k^{*},$$
the test that rejects $H_0$ for large values of $Q$ is equivalent to the test that rejects $H_0$ for large values of $Q^{*}$ because $Z_i^{*}$ is a monotone function of $Z_{i}$. Letting $\*Z^{*}=(Z_{n-m_1}^{*},\ldots,Z_{n-m_0}^{*})$, we approximate $\*Z^{*}$ with a multivariate normal distribution with mean $\*0$ and covariance $\*\Sigma$, where $\*\Sigma$ can be empirically approximated or obtained analytically through a Taylor series expansion of $Z_i^{*}$. Note that although each $Z_i^*$ are marginally normally distributed, the dependence structure and nonlinear transformation used to obtain the $Z_i^*$ makes the use of the multivariate normal for their joint distribution only an approximation. Despite this, next we will demonstrate that the multivariate normal approximation is quite accurate.

The first-order Taylor series approximation of $\mbox{cor}(Z_{j_1}^{*},Z_{j_2}^{*})$ is $(n-j_1)/(n-j_2)$ for $j_1 < j_2$. Once $\*\Sigma$ is estimated, numerical integration on the $MVN(\*0,\hat{\*\Sigma})$ distribution can be used to obtain:
$$\mbox{pr}(Q\geq b^2) =  1-\mbox{pr}\left[Q^{*} < \Phi^{-1}\{F_q(b^2)\}\right].$$
As is demonstrated in Section \ref{sec:pvalacc}, empirical estimation of $\*\Sigma$ is an easier task than direct empirical estimation of  $\mbox{pr}(Q\geq b^2)$, as inference of $Q$ is fairly robust to the estimation of $\*\Sigma$, so fewer simulations are required for accurate p-values than would be needed for Monte Carlo simulation of the null distribution directly.

\begin{table}
\caption{Type I error simulations for likelihood ratio tests. The asymptotic p-values are generated according to equation \ref{pvalasymp}. The other two approaches obtain p-values using the $\*Z^{*}$ with its covariance $\*\Sigma$ estimated empirically or through a 1st order approximation. The $0. 05$ significance level is used with $10000$ simulations for each setting.}
\begin{center}
\begin{tabular}{cc|ccc|ccc}
 \\
&& \multicolumn{3}{c}{$m_1=6$ }& \multicolumn{3}{c}{$m_1=\sqrt{n}$} \\
&& \multicolumn{3}{c}{$m_0=0$ }& \multicolumn{3}{c}{$m_0=\sqrt{n}/2$} \\
q & n & Asymptotic & Approx. $\hat{\*\Sigma}$  & Emp. $\hat{\*\Sigma}$ & Asymptotic & Approx. $\hat{\*\Sigma}$ & Emp. $\hat{\*\Sigma}$ \\[5pt]
5 & 30 & $0.136$ & $0.034$ &$0.044$ & $0.082$ & $0.051$ & $0.039$ \\
5 & 100 & $0.133$ &$0.036$ &$0.041$ & $0.070$ & $0.089$ & $0.037$ \\
5 & 365 & $0.144$&$0.035$ & $0.041$ & $0.068$ & $0.141$ & $0.039$ \\
5 & 1000  & $0.141$ & $0.042$ & $0.045$ & $0.055$&$0.173$ & $0.035$ \\
10 & 30 & $0.169$ & $0.030$ &$0.042$ & $0.095$ & $0.070$ & $0.044$ \\
10 & 100 & $0.175$ &$0.041$ &$0.045$ & $0.085$ & $0.097$ & $0.040$ \\
10 & 365 & $0.340$&$0.048$ & $0.048$ & $0.077$ & $0.150$ & $0.042$ \\
10 & 1000  & $0.163$ & $0.040$ & $0.042$ & $0.069$&$0.185$ & $0.043$ \\
50 & 30 & $0.332$&$0.038$ &$0.045$ & $0.197$ & $0.079$ & $0.050$ \\
50 & 100 & $0.329$ &$0.049$ & $0.050$ & $0.169$ & $0.101$ & $0.047$ \\
50 & 365 & $0.340$ & $0.048$ & $0.049$ & $0.154$ & $0.151$ & $0.047$ \\
50 & 1000 & $0.340$ & $0.050$ & $0.048$ & $0.138$&$0.188$ & $0.047$   \\
\end{tabular}
\end{center}
\label{tablet1error}
\end{table}

To demonstrate the inaccuracy of asymptotic p-values in settings that mirror most digital phenotyping studies, we simulate according to model \ref{model} under $H_0$ with $\delta_i=0$ and with $n=30, 100, 365, \mbox{ and } 1000$, representings days of data collected, and $q=5, 10 \mbox{ and } 50$, representing the number of features collected on each day. Three methods for obtaining p-values are compared: asymptotic p-values according to equation \ref{pvalasymp}, p-values based on $Q^{*}$ using numerical integration of the cumulative multivariate normal distribution where $\*\Sigma$ is estimated empirically, and the same approach with $\*\Sigma$ estimated using its first order approximation given by equation \ref{1storder}. Table~\ref{tablet1error} demonstrates the Type I error at the $0.05$ significance level for these methods in a variety of settings. The $m_0=0$ and $m_1=6$ setting corresponds to digital phenotyping studies which aim to identify change points in the most recent week, while the $m_0 = \sqrt{n}/2$ and $m_1=\sqrt{n}$ setting mirrors traditional asymptotic change point detection treatments. In the $m_0 = \sqrt{n}/2$ and $m_1=\sqrt{n}$ setting, the asymptotic approximation is much improved, particularly for large $n$ and small $q$. In the $m_0=0$ and $m_1=6$ setting, the asymptotic p-values will remain inaccurate even with very large $n$ because all asymptotic treatments of change point detection problems require both $m_0$ and $m_1$ to increase as well. In contrast, the two methods that rely on estimation of $\*\Sigma$ are more accurate when $m_1=6$ and $m_0=0$ than when $m_1=\sqrt{n}$ and $m_0=\sqrt{n}/2$ because the dimension of $\*\Sigma$ is $m^* \times m^*$ where $m^* = m_1-m_0+1$, making estimation easier when $m^*$ is small. Overall, empirical estimation of $\*\Sigma$ is the most accurate approach, with slightly conservative performance due to the multivariate normal approximation of its joint distribution, with the first order approximation of $\*\Sigma$ also having an acceptable albeit more conservative performance in the $m_1=6$ and $m_0=0$ case. Due to its high accuracy, the method that employs empirical estimation of $\*\Sigma$ will be used to perform inference for the likelihood ratio test in the remaining power simulations and data analysis sections.

\section{Accuracy of empirically-based p-values}
\label{sec:pvalacc}

\begin{figure}
\begin{center}
\includegraphics[width=5in]{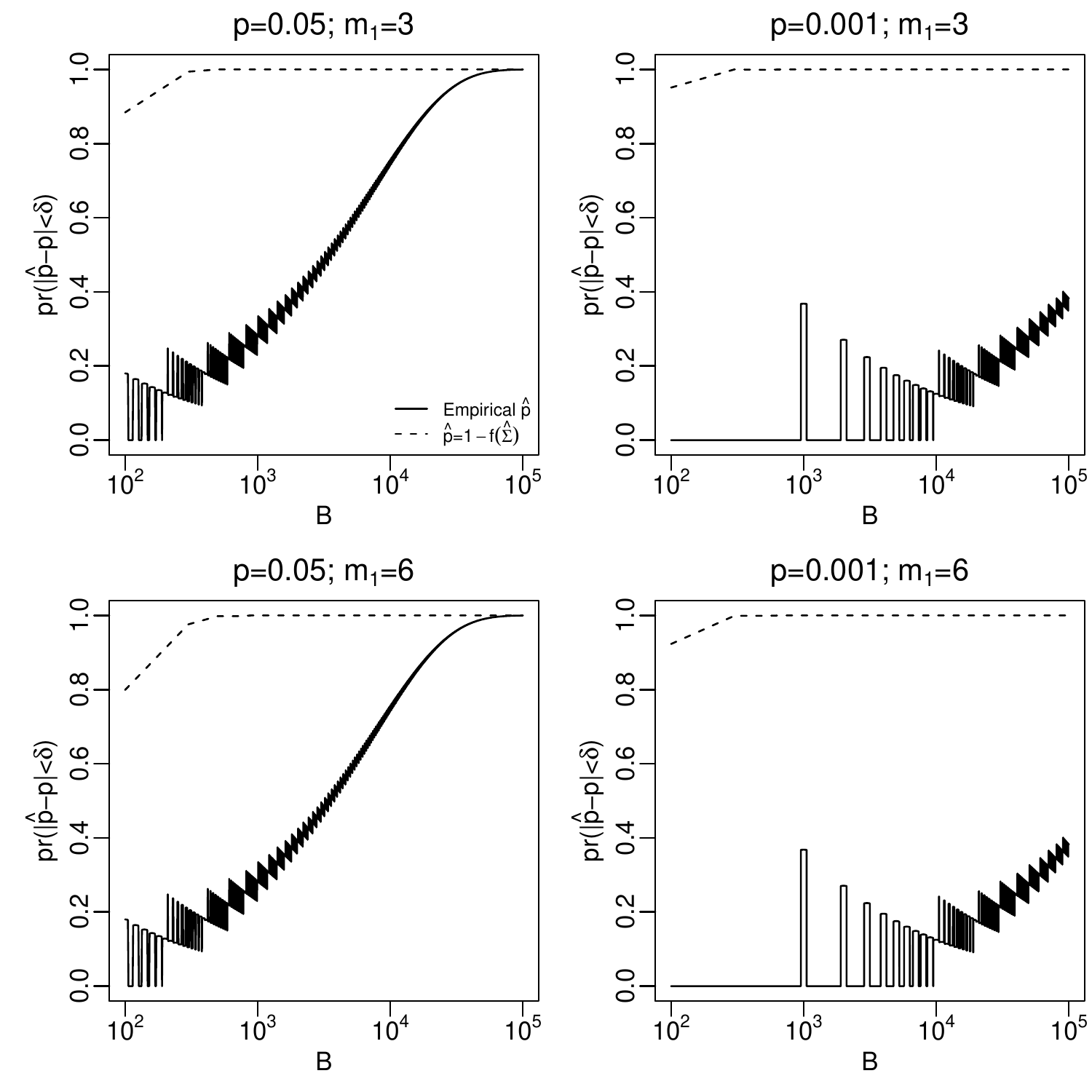}
\end{center}
\caption{ Accuracy of estimated p-values. The accuracy of binomial proportion empirical estimates of $p$, $\tilde{p}$ (solid lines), are compared with estimates of $p$ that are obtained from integrating over the multivariate normal density using an empirical estimator of $\*\Sigma$, $\hat{p}$ (dotted lines). In both cases, $B$ repeated simulations under $H_0$ are used to estimate $p$. Accuracy is determined by the fraction of estimated p-values that are within $5\%$ of $p$ (i.e. $\delta=0.05 p$). In all cases $m_0=0$, representing online change point detection scenarios.}
\label{fig1}
\end{figure}

For a given critical value $a$  of the test statistic $Q^*$, let the corresponding p-value be $p = \mbox{pr}(Q^* \geq a) = 1 - f(\*\Sigma)$, where 
$$f(\*\Sigma) = \frac{1}{\sqrt{|\*\Sigma|(2\pi)^{m^*}}}\int_{-\infty}^{a}\cdots \int_{-\infty}^{a}\exp\left(-\frac{1}{2}\*y^T\*\Sigma^{-1}\*y\right)d\*y.$$
As $\*\Sigma$ is unobserved, to compute this integral we must estimate $\*\Sigma$ instead. Suppose we generate $\*Z^{*(b)}$ for $b = 1,\ldots,B$, each under $H_0$. Letting $\hat{\*\Sigma}$ be empirical correlation of the $\*Z^{*(b)}$, in Section \ref{sec:methodsp} we proposed using
$$\hat{p} = 1- f\left(\hat{\*\Sigma}\right)$$
to estimate the p-value. This is an alternative to directly estimating the p-value empirically using
$$\tilde{p} = \sum_{b=1}^B I_{\{Q^{*(b)}\geq a\}}/B.$$
The variance of  $\tilde{p}$ is $p(1-p)/B$ as it is a binomial proportion. For comparison, we find the variance of $\hat{p}$:
\begin{theorem}
\label{empdeltap}
Given that  $m^* = m_1-m_0+1$ is the length of $\*Z^*$ and let $\hat{\*\Sigma}$ be the empirical estimator of $\*\Sigma$ based on $B$ simulated instances of $\*Z^*$ under $H_0$. Letting $\hat{p} = 1-f\left(\hat{\*\Sigma}\right)$, then:
\begin{equation}
\label{varphat}
\mbox{var}(\hat{p}) = c_a \cdot \frac{p^2}{B} + o_B\left(B^{-1}\right) 
\end{equation}
where 
\[
\begin{array}{rl}
c_a =&\sum\limits_{1 \leq i<j \leq m^{*}}\left\{\left(\*\Sigma^{-1}\right)_{ij}-h_{ij}(a)\right\}^2\left(1-\Sigma_{ij}^2\right)^2 \\
& + 2\sum\limits_{1 \leq i<j<k \leq m^{*}} \left\{\left(\*\Sigma^{-1}\right)_{ij}-h_{ij}(a)\right\} \left\{\left(\*\Sigma^{-1}\right)_{jk}-h_{jk}(a)\right\}(\Sigma_{ik} + \Sigma_{ij}\Sigma_{jk})
\end{array}
\]
and
\[
\begin{array}{rl}
h_{ij}(a) &=\frac{1}{1-f(\*\Sigma)}\displaystyle\int\limits_{\*y: \max_j{y_j}>a}\frac{1}{2}\*y^T  \*\Sigma^{-1}\frac{\delta \*\Sigma}{\delta \Sigma_{ij}}\*\Sigma^{-1} \*y \frac{\exp\left(-\frac{1}{2}\*y^T\*\Sigma^{-1}\*y\right)}{ |\*\Sigma|^{1/2}(2\pi)^{m^*/2}}d\*y \\
&= E\left(\frac{1}{2}\*y^T  \*\Sigma^{-1}\frac{\delta \*\Sigma}{\delta \Sigma_{ij}}\*\Sigma^{-1} \*y\bigg|\max_j{y_j}>a\right). \\
\end{array}
\]
\end{theorem}

While the $c_a$ term in Equation \ref{varphat} is an increasing function of $p$ as $p\rightarrow 0$, it is increasing at most at the rate of $(\log p )^2$ and so $\mbox{var}(\hat{p})$ is dominated by the $p^2$ term. The following corollary to Theorem \ref{empdeltap} demonstrates this in the direct comparison of the variances of $\hat{p}$ and $\tilde{p}$.

\begin{corollary}\label{varcompare}
There exists $c^*$ such that for all $p<0.5$ we have the following relation:
\begin{equation}\label{varcompareeq}
\frac{\mbox{var}(\hat{p})}{\mbox{var}(\tilde{p})} \leq c^*\cdot \frac{p(\log p)^2}{(1-p)} + o_B(1)
\end{equation}
\end{corollary}

The implication of Equation \ref{varcompareeq} is that $\hat{p}$ is a far superior estimator than $\tilde{p}$ for reasonably small p-values. However, in situations where $c_a$ is large and $p$ is not small, $\tilde{p}$ can be a more efficient p-value estimator than $\hat{p}$. Because $\*\Sigma$ is $m^* \times m^*$, when $m^*$ is large the empirical estimator $\hat{\*\Sigma}$ is noisy, which translates to a larger $c_a$ and higher variability in $\hat{p}$.

A good p-value estimator should have a high likelihood of being close to the true $p$. As a more interpretable alternative to looking at variability, we compare $\mbox{pr}(|\hat{p}-p| < \delta)$ to $\mbox{pr}(|\tilde{p}-p| < \delta)$ through a simulation study, where we let $p=0.05,  0.001$, and for a given $B$ used $1000$ iterations to determine the fraction of the time that both $\hat{p}$ and $\tilde{p}$ were within $5\%$  of $p$ (i.e. $\delta = 0.05p$). The results of these simulations are shown in Figure \ref{fig1}. The discrete nature of a binomial proportion of a rare event explain the erratic nature of $\tilde{p}$. It is clear that a minimum of $B=p^{-1}$ iterations are required for $\tilde{p}$ to achieve the correct p-value, and even then the likelihood of doing so is quite small. Only when $B$ is many orders of magnitude larger than $p^{-1}$ does $\tilde{p}$ become a useful estimator for $p$. In contrast, $\hat{p}$ is a reliable estimator of $p$ even for very small $B$. Increasing $m^*$ by going from $m_1=3$ to $m_1=6$ does reduce the accuracy of $\hat{p}$ by increasing $c_a$, but the reduction in  $\mbox{pr}(|\hat{p}-p| < \delta)$ is relatively small and only relevant for small $B$. Additionally, the performance of $\hat{p}$ is robust to the magnitude of $p$, making it an ideal estimator for moderately small to very small $p$.

\section{Numerical power comparison}
\label{sec:power}

We also consider several alternatives to the likelihood ratio test for multivariate change point detection problems designed for testing model \ref{model}. Hotelling's multivariate $T^2$ tests for a change point on the $j$th day using the statistic $\sum_{i=1}^q \hat{\epsilon}_{ij}^2$, where $\hat{\epsilon}_{ij}$ is estimated from model \ref{model} under $H_0$   \citep{hotelling1947multivariate}. The multivariate extension of the cumulative control sum chart procedure proposed by \citet{crosier1988multivariate} sets a target value $a>0$ and lets $C_j = \{\sum_{i=1}^q (s_{ij}+\hat{\epsilon}_{ij}-a)^2\}^{1/2}$ where $s_{ij}=0 \; (i=1,\ldots,q)$ when $C_j\leq \kappa$ for some threshold $\kappa>0$, and $s_{ij}=(s_{i(j-1)}+\hat{\epsilon}_{ij}-a)(1-\kappa/C_j)$ otherwise. This process is initialized so that $s_{i0}=0 \; (i=1,\ldots,q)$, and $H_0$ is rejected for large values of $\max_j(\sum_{i=1}^q s_{ij}^2)$. This test can be extended to two-sided hypotheses by negating the sign of each $\hat{\epsilon}_{ij}$ and repeating the procedure. This test has similarities to Hotelling's $T^2$, but in contrast, it is designed to cumulate changes in the mean across multiple consecutive timepoints.

\begin{figure}
\begin{center}
\includegraphics[width=5in]{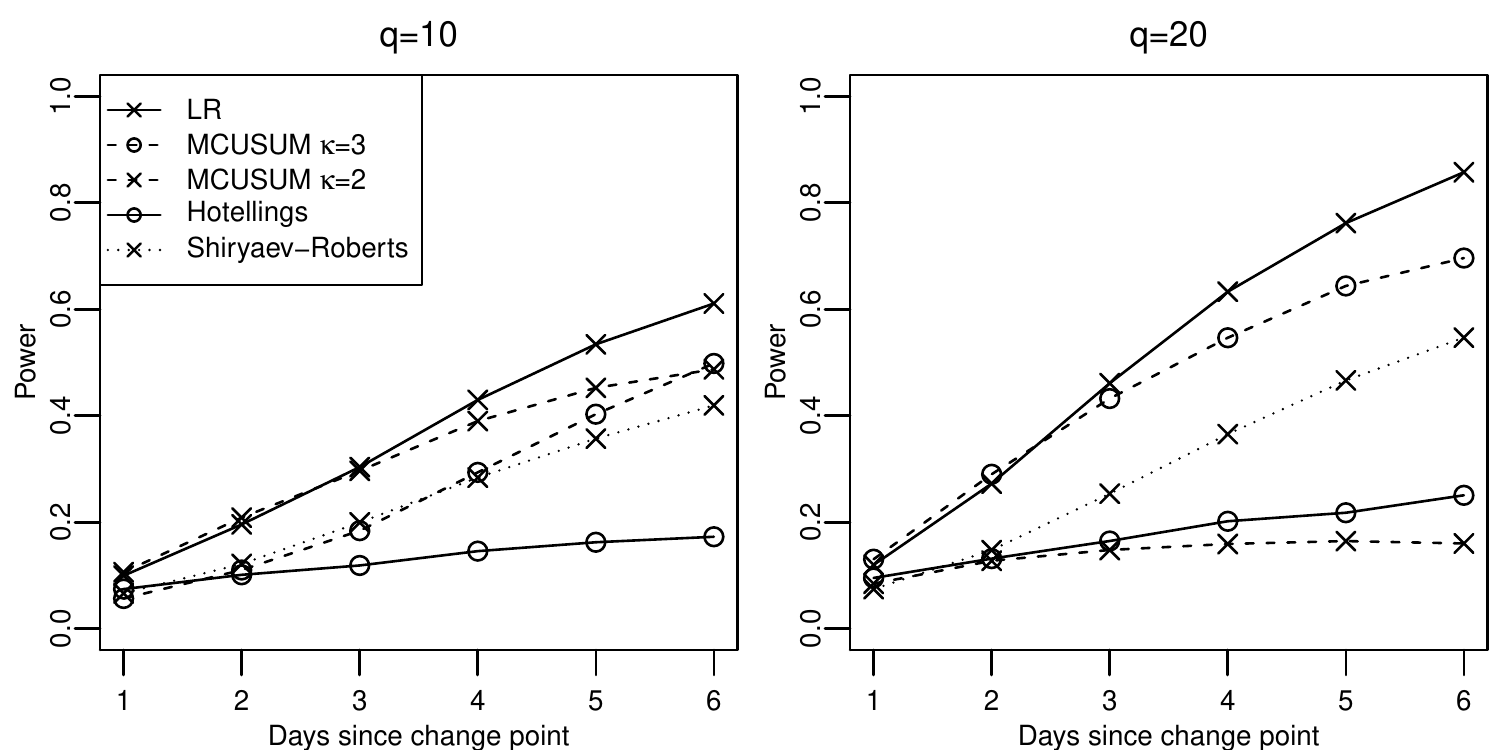}
\end{center}
\caption{A power comparison of multivariate change point detection methods. With the change point location $k$ ranging from $n-1$ to $n-6$, where $n=100$, this simulation corresponds to change point detection in an individual after $100$ days of digital phenotyping monitoring, with change points occuring at various points in the past week. Features are generated independently according to model \ref{model} with each $\delta_i=0.5$. }
\label{fig2}
\end{figure}

To compare the performance of these methods in non-asymptotic settings meant to mirror digital phenotyping studies, we consider $n=100$ days of observation with the change point location $k$ occuring in the week prior to the end of followup (day $n$). The number of features considered were $q=10$ and $q=20$, and data was simulated according to model \ref{model} where $\delta_i=0.5$. Two different thresholds were used for the multivariate cumulative control sum chart procedure, $\kappa=2$ and $\kappa=3$. Power was estimated for each parameter setting and each change point location from 1000 iterations, and the results are in Figure \ref{fig2}.

Consistently the likelihood ratio test had the highest power, followed by multivariate cumulative control sum chart procedure, and then Hotelling's $T^2$. Each method gains power as the change point shifted further back in time, which intuitively correponds to increasing the number of observations post-change point. The multivariate cumulative sum control chart procedure benefits from using a smaller threshold $\kappa$  as the number of features increase. Increasing the number of features from $q=10$ to $q=20$ consistently increased the power of all methods, and demonstrates the necessity of multivariate and diverse feature sets in non-asymptotic change point problems. With so few observations after the change point the only feasible way of detecting a change point is if it is observed simultaneously across a variety of features. Univariate change point detection would be severely underpowered in these settings.

\section{Change-point detection to predict relapse}
\label{sec:realdata}

\begin{figure}
\begin{center}
\includegraphics{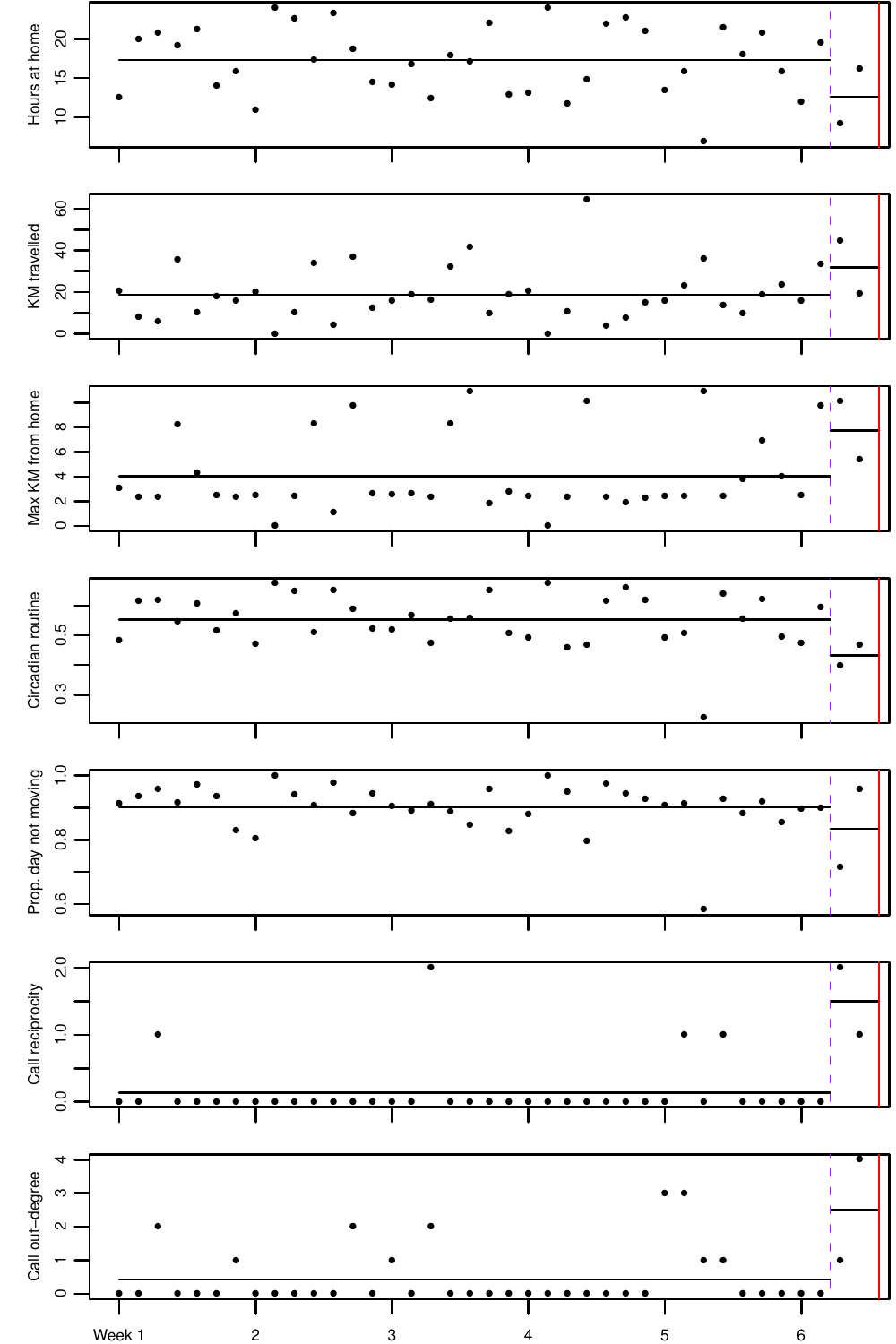}
\caption{Changes in behavioral features prior to relapse in a patient with schizophrenia. The vertical dotted line represents the time of the change point and the vertical solid line represents the time of relapse and hospitalization. The horizontal solid lines represents each feature means before and after the change point.}
\label{fig3}
\end{center}
\end{figure}

We consider the daily behavioral features of a patient with schizophrenia as part of a digital phenotyping pilot study that recruited subjects from a state mental health clinic in Boston, Massachusetts. During recruitment, the patient installed the Beiwe application on their personal smartphone, and during follow up GPS, accelerometer, call logs, and other data streams were collected through background usage \citep{torous2016new}. We consider a set of seven features representing daily summaries of patient behavior related to mobility, activity, and sociability: hours spent at home, distance travelled (km), maximum distance from home (km), proportion of the day not moving, call out-degree, call reciprocity, and circadian routine. Circadian routine is quantified from daily GPS data as detailed in \citet{canzian2015trajectories}.

After six weeks of follow up, the patient relapsed and was hospitalized. Here we retrospectively employ multivariate change point detection methods on these features to determine if relapse could have been anticipated based on changes in behavior leading up to the patient's hospitalization. By considering only the 39 days of data that were observed prior to relapse, each change point detection method tested the previous week leading up to relapse for the occurence of a change point in the features. Hotelling's $T^2$ as well as the multivariate cumulative sum control chart procedure with both $\kappa=2$ and $\kappa=3$ were unable to detect a significant change point in the features, but the likelihood ratio test detected a change in behavior two days before relapse with marginal significance ($p=0.041$).

A visual representation of the behavioral features and their change prior to relapse are demonstrated in Figure \ref{fig3}. Two days prior to relapse there was a marked increase in the number of contacts the patient placed calls to, an increase in the rate of returning calls, a sharp departure from routine, more time spent out of the home, and trips to places further away from home than usual. While the departures from the mean for each feature alone are not necessarily noteworthy, together their simultaneous change speaks to a broader pattern that enables detection based on their combined signals. While this result is promising, in practice marginal significance will not be sufficient due to multiple comparison corrections. Power can most reliably be gained by augmenting the digital phenotyping feature sets to better quantify behavior.

\section{Discussion}
\label{sec:discussion}

For non-asymptotic univariate change point detection, very strong signals are required to identify changes from a relatively few number of observations. For this reason, multivariate change point detection is capable of aggregating information across features so as to detect change points even when each feature's marginal change is moderately small. The seven behavioral features represented in Section \ref{sec:realdata}  are not exhaustive and do not fully represent the useful information that can be extracted from smartphone digital phenotyping data. As digital phenotyping studies move beyond feasibility and pilot studies, standardized methods and software will need to be developed for data cleaning and quantification into a more comprehensive set of meaningful behavioral features relating to mobility, sleep, activity, and sobiability. As $q$ increases, it may be necessary to apply dimension reduction to alleviate redundancies in the feature set while increasing power.

While the data example presented in Section \ref{sec:realdata} is a retrospective analysis, it is ultimately necessary for change point detection methods to be applied in an online fashion so as to detect changes in behavior before adverse events occur so as to prompt timely interventions. Many mental health outcomes of interest in digital phenotyping studies are rare, such as suicide attempts or relapse, and so new large-scale digital phenotyping studies are needed to investigate the sensitivity and specificity of multivariate change point detection methods. This will help identify the appropriate significance levels to use in practice so as to both minimize the frequency of false positives while still remaining sensitive enough to detect subtle behavioral changes. Indeed, while multivariate change point detection is a vital first step, there is a lack of consensus on what constitutes the ideal intervention or course of action to take after a change point is detected.


\backmatter


\section*{Supplementary Materials}

The schizophrenia digital phenotypes used in Section~\ref{sec:realdata} as well as the code to reproduce all results is available with
this paper at the Biometrics website on Wiley Online
Library.\vspace*{-8pt}


%
  \bibliographystyle{biom} 
 \bibliography{paper-ref}

\begin{thebibliography}{}

\bibitem[\protect\citeauthoryear{Bardwell, Fearnhead, Eckley, Smith, and
  Spott}{Bardwell et~al.}{2018}]{bardwell2018most}
Bardwell, L., Fearnhead, P., Eckley, I.~A., Smith, S., and Spott, M. (2018).
\newblock Most recent changepoint detection in panel data.
\newblock {\em Technometrics} pages 1--28.

\bibitem[\protect\citeauthoryear{Billingsley}{Billingsley}{2008}]{billingsley2008probability}
Billingsley, P. (2008).
\newblock {\em Probability and measure}.
\newblock John Wiley \& Sons.

\bibitem[\protect\citeauthoryear{Canzian and Musolesi}{Canzian and
  Musolesi}{2015}]{canzian2015trajectories}
Canzian, L. and Musolesi, M. (2015).
\newblock Trajectories of depression: unobtrusive monitoring of depressive
  states by means of smartphone mobility traces analysis.
\newblock In {\em Proceedings of the 2015 ACM international joint conference on
  pervasive and ubiquitous computing}, pages 1293--1304. ACM.

\bibitem[\protect\citeauthoryear{Cho et~al\mbox{.}}{Cho
  et~al.}{2016}]{cho2016change}
Cho, H. et~al. (2016).
\newblock Change-point detection in panel data via double cusum statistic.
\newblock {\em Electronic Journal of Statistics} {\bf 10,} 2000--2038.

\bibitem[\protect\citeauthoryear{Cho and Fryzlewicz}{Cho and
  Fryzlewicz}{2015}]{cho2015multiple}
Cho, H. and Fryzlewicz, P. (2015).
\newblock Multiple-change-point detection for high dimensional time series via
  sparsified binary segmentation.
\newblock {\em Journal of the Royal Statistical Society: Series B (Statistical
  Methodology)} {\bf 77,} 475--507.

\bibitem[\protect\citeauthoryear{Crosier}{Crosier}{1988}]{crosier1988multivariate}
Crosier, R.~B. (1988).
\newblock Multivariate generalizations of cumulative sum quality-control
  schemes.
\newblock {\em Technometrics} {\bf 30,} 291--303.

\bibitem[\protect\citeauthoryear{Hogan and Siegmund}{Hogan and
  Siegmund}{1986}]{hogan1986large}
Hogan, M. and Siegmund, D. (1986).
\newblock Large deviations for the maxima of some random fields.
\newblock {\em Advances in Applied Mathematics} {\bf 7,} 2--22.

\bibitem[\protect\citeauthoryear{Hotelling}{Hotelling}{1947}]{hotelling1947multivariate}
Hotelling, H. (1947).
\newblock Multivariate quality control.
\newblock {\em Techniques of statistical analysis} .

\bibitem[\protect\citeauthoryear{Hotelling}{Hotelling}{1953}]{hotelling1953new}
Hotelling, H. (1953).
\newblock New light on the correlation coefficient and its transforms.
\newblock {\em Journal of the Royal Statistical Society. Series B
  (Methodological)} {\bf 15,} 193--232.

\bibitem[\protect\citeauthoryear{Insel}{Insel}{2017}]{insel2017digital}
Insel, T.~R. (2017).
\newblock Digital phenotyping: Technology for a new science of behavior.
\newblock {\em Jama} {\bf 318,} 1215--1216.

\bibitem[\protect\citeauthoryear{James, James, and Siegmund}{James
  et~al.}{1987}]{james1987tests}
James, B., James, K.~L., and Siegmund, D. (1987).
\newblock Tests for a change-point.
\newblock {\em Biometrika} {\bf 74,} 71--83.

\bibitem[\protect\citeauthoryear{Kiefer}{Kiefer}{1959}]{kiefer1959k}
Kiefer, J. (1959).
\newblock K-sample analogues of the kolmogorov-smirnov and cram{\'e}r-v. mises
  tests.
\newblock {\em The Annals of Mathematical Statistics} pages 420--447.

\bibitem[\protect\citeauthoryear{Killick, Fearnhead, and Eckley}{Killick
  et~al.}{2012}]{killick2012optimal}
Killick, R., Fearnhead, P., and Eckley, I.~A. (2012).
\newblock Optimal detection of changepoints with a linear computational cost.
\newblock {\em Journal of the American Statistical Association} {\bf 107,}
  1590--1598.

\bibitem[\protect\citeauthoryear{Kirch, Muhsal, and Ombao}{Kirch
  et~al.}{2015}]{kirch2015detection}
Kirch, C., Muhsal, B., and Ombao, H. (2015).
\newblock Detection of changes in multivariate time series with application to
  eeg data.
\newblock {\em Journal of the American Statistical Association} {\bf 110,}
  1197--1216.

\bibitem[\protect\citeauthoryear{Lavielle and Teyssiere}{Lavielle and
  Teyssiere}{2006}]{lavielle2006detection}
Lavielle, M. and Teyssiere, G. (2006).
\newblock Detection of multiple change-points in multivariate time series.
\newblock {\em Lithuanian Mathematical Journal} {\bf 46,} 287--306.

\bibitem[\protect\citeauthoryear{Ma and Yau}{Ma and Yau}{2016}]{ma2016pairwise}
Ma, T.~F. and Yau, C.~Y. (2016).
\newblock A pairwise likelihood-based approach for changepoint detection in
  multivariate time series models.
\newblock {\em Biometrika} {\bf 103,} 409--421.

\bibitem[\protect\citeauthoryear{Matteson and James}{Matteson and
  James}{2014}]{matteson2014nonparametric}
Matteson, D.~S. and James, N.~A. (2014).
\newblock A nonparametric approach for multiple change point analysis of
  multivariate data.
\newblock {\em Journal of the American Statistical Association} {\bf 109,}
  334--345.

\bibitem[\protect\citeauthoryear{Nardi, Siegmund, Yakir, et~al\mbox{.}}{Nardi
  et~al.}{2008}]{nardi2008distribution}
Nardi, Y., Siegmund, D.~O., Yakir, B., et~al. (2008).
\newblock The distribution of maxima of approximately gaussian random fields.
\newblock {\em The Annals of Statistics} {\bf 36,} 1375--1403.

\bibitem[\protect\citeauthoryear{Olshen, Venkatraman, Lucito, and
  Wigler}{Olshen et~al.}{2004}]{olshen2004circular}
Olshen, A.~B., Venkatraman, E., Lucito, R., and Wigler, M. (2004).
\newblock Circular binary segmentation for the analysis of array-based dna copy
  number data.
\newblock {\em Biostatistics} {\bf 5,} 557--572.

\bibitem[\protect\citeauthoryear{Onnela and Rauch}{Onnela and
  Rauch}{2016}]{onnela2016harnessing}
Onnela, J.-P. and Rauch, S.~L. (2016).
\newblock Harnessing smartphone-based digital phenotyping to enhance behavioral
  and mental health.
\newblock {\em Neuropsychopharmacology: official publication of the American
  College of Neuropsychopharmacology} {\bf 41,} 1691--1696.

\bibitem[\protect\citeauthoryear{Page}{Page}{1954}]{page1954continuous}
Page, E.~S. (1954).
\newblock Continuous inspection schemes.
\newblock {\em Biometrika} {\bf 41,} 100--115.

\bibitem[\protect\citeauthoryear{Preuss, Puchstein, and Dette}{Preuss
  et~al.}{2015}]{preuss2015detection}
Preuss, P., Puchstein, R., and Dette, H. (2015).
\newblock Detection of multiple structural breaks in multivariate time series.
\newblock {\em Journal of the American Statistical Association} {\bf 110,}
  654--668.

\bibitem[\protect\citeauthoryear{Siegmund}{Siegmund}{2013}]{siegmund2013sequential}
Siegmund, D. (2013).
\newblock {\em Sequential analysis: tests and confidence intervals}.
\newblock Springer Science \& Business Media.

\bibitem[\protect\citeauthoryear{Siegmund, Yakir, and Zhang}{Siegmund
  et~al.}{2011}]{siegmund2011detecting}
Siegmund, D., Yakir, B., and Zhang, N.~R. (2011).
\newblock Detecting simultaneous variant intervals in aligned sequences.
\newblock {\em The Annals of Applied Statistics} pages 645--668.

\bibitem[\protect\citeauthoryear{Siegmund}{Siegmund}{1992}]{siegmund1992tail}
Siegmund, D.~O. (1992).
\newblock Tail approximations for maxima of random fields.
\newblock In {\em Probability Theory: Proceedings of the 1989 Singapore
  Probability Conference held at the National University of Singapore, June
  8--16, 1989}, page 147. Walter de Gruyter GmbH \& Co KG.

\bibitem[\protect\citeauthoryear{Torous, Kiang, Lorme, and Onnela}{Torous
  et~al.}{2016}]{torous2016new}
Torous, J., Kiang, M.~V., Lorme, J., and Onnela, J.-P. (2016).
\newblock New tools for new research in psychiatry: a scalable and customizable
  platform to empower data driven smartphone research.
\newblock {\em JMIR mental health} {\bf 3,}.

\bibitem[\protect\citeauthoryear{Torous, Onnela, and Keshavan}{Torous
  et~al.}{2017}]{torous2017new}
Torous, J., Onnela, J., and Keshavan, M. (2017).
\newblock New dimensions and new tools to realize the potential of rdoc:
  digital phenotyping via smartphones and connected devices.
\newblock {\em Translational psychiatry} {\bf 7,} e1053.

\bibitem[\protect\citeauthoryear{Woodroofe}{Woodroofe}{1982}]{woodroofe1982nonlinear}
Woodroofe, M. (1982).
\newblock {\em Nonlinear renewal theory in sequential analysis}.
\newblock SIAM.

\bibitem[\protect\citeauthoryear{Xie, Huang, and Willett}{Xie
  et~al.}{2013}]{xie2013change}
Xie, Y., Huang, J., and Willett, R. (2013).
\newblock Change-point detection for high-dimensional time series with missing
  data.
\newblock {\em IEEE Journal of Selected Topics in Signal Processing} {\bf 7,}
  12--27.

\bibitem[\protect\citeauthoryear{Zhang, Siegmund, Ji, and Li}{Zhang
  et~al.}{2010}]{zhang2010detecting}
Zhang, N.~R., Siegmund, D.~O., Ji, H., and Li, J.~Z. (2010).
\newblock Detecting simultaneous changepoints in multiple sequences.
\newblock {\em Biometrika} {\bf 97,} 631--645.

\end{thebibliography}

%
%
%

\appendix


\section{}

\subsection{Proof of Theorem 1}

\begin{proof}
For $k\geq n-m_1$, $U_{ik} = \sum_{j=k+1}^nY_{ij}/\sqrt{n-k} + o(1)$. Let $P_\delta$ represent the probability measure for  $\sqrt{k}(U_{1k},\dots,U_{qk})^T$ where $E(Y_{ij})=\delta_i$. We use the approach of starting with the likelihood ratio identity commonly used in similar asymptotic tail probability estimations \cite{nardi2008distribution}.The likelihood ratio of $P_0$ with respect to $A = \prod_{i=1}^q\int_{-\infty}^{\infty}P_{\delta_i}\phi(\delta_i) d\delta_i$ is  

\[
\begin{array}{rl}
l(k;A,P_0) &=  \int_{-\infty}^{\infty}\cdots \int_{-\infty}^{\infty}\prod_{i=1}^q\frac{\frac{1}{\sqrt{2\pi k}}\exp\left\{-\left(\sum_{j=k+1}^nY_{ij}-(n-k)\delta_i\right)^2/2(n-k)\right\}}{\frac{1}{\sqrt{2\pi k}}\exp\left\{-\left(\sum_{j=k+1}^nY_{ij}\right)^2/2(n-k)\right\}}\phi(\delta_i)d\delta_i \\
			&= (n-k+1)^{-q/2}\exp\{\sum_{i=1}^q(\sum_{j=k+1}^nY_{ij})^2/2(n-k+1)\}\\
\end{array}
\]
Letting $K =\max(k;n-m_1\leq k\leq n-m_0, Z_k\geq b^2)$, Wald's likelihood ratio identity gives

\[
\begin{array}{rl}
\mbox{pr}(Q\geq b^2;P_0) &= E_A\left\{l(K;P_0,A)I_{\{Q\geq b\}}\right\} \\
							&=  \int_{-\infty}^{\infty}\cdots \int_{-\infty}^{\infty}(n-K+1)^{q/2}\exp\{-\sum_{i=1}^q(\sum_{j=K+1}^nY_{ij})^2/2(n-K+1)\} \\
									&\;\;\;\;\cdot I_{\{Q\geq b^2\}}\prod_{i=1}^q\phi(\delta_i)d\delta_i\\
\end{array}
\]

Because $K$ is the index where $Z_k$ crosses over $b^2$,  we have $Z_K = b^2 +o(1)$, which, generalizing from 4.12 of \cite{siegmund2013sequential}, implies  $\mbox{pr}\{\lim_{b\rightarrow\infty}(n-K)b^{-2}=(\sum_{i=1}^q\delta_i^2)^{-1}\}$. Thus, $Q \geq b^2$ is asymptotically equivalent to $c_1 \leq \theta  \leq c_0$, where $\theta=(\sum_{i=1}^q \delta_i^2)^{-1/2}$.

\[
\begin{array}{rl}
\mbox{pr}(Q\geq b^2;P_0) &=  \int_{-\infty}^{\infty}\cdots \int_{-\infty}^{\infty}(n-K+1)^{q/2}\exp\{-(n-K)b^2/2(n-K+1)\} \\
									&\;\;\;\;\cdot I_{\{c_1 \leq \theta \leq c_0\}}\prod_{i=1}^q\phi(\delta_i)d\delta_i\\
							    &=  (2\pi)^{-q/2}b^qexp(-b^2/2)\int_{-\infty}^{\infty}\cdots \int_{-\infty}^{\infty}\theta^{-q} I_{\{c_1 \leq \theta \leq c_0\}}d\delta_1\cdots d\delta_q\\
							    &=  (2\pi)^{-q/2}b^qexp(-b^2/2)\int_{c_1}^{c_0}\theta^{-q}\left\{\frac{2\pi^{q/2}\theta^{q-1}}{\Gamma(q/2)}\right\} d\theta\\
\end{array}
\]

This last equality results from observing that the $q$-dimensional integral is over the surface area of spheres in $\mathbb{R}^q$, or $(2\pi^{q/2}\theta^{q-1})/\Gamma(q/2)$, with radii ranging from $c_1 \leq \theta \leq c_0$. Observing that $\int_{c_1}^{c_0}\theta^{-1}d\theta = 2\ln(c_0/c_1) = \ln(m_1/m_0)$ leads to \ref{pvalasymp}.
\end{proof}

\subsection{Proof of Theorem 2}

\begin{proof}
By expanding $f\left(\hat{\*\Sigma}\right)$ around $\*\Sigma$, a symmetric matrix, we obtain:
$$ \sqrt{B}f\left(\hat{\*\Sigma}\right) = \sqrt{B}f(\*\Sigma) + \sqrt{B} \sum_{i<j} \frac{\delta f(\*\Sigma)}{\delta \Sigma_{ij}} (\hat{\Sigma}_{ij} - \Sigma_{ij}) + o_B(1).$$
Differentiating $f(\*\Sigma)$ with respect to $\Sigma_{ij}$ yields
\[
\begin{array}{rl}
\frac{\delta f\left(\*\Sigma\right)}{\delta \Sigma_{ij}} =& -\frac{1}{2}|\*\Sigma|^{-3/2}(2\pi)^{-m^*/2}\frac{\delta |\*\Sigma|}{\delta \Sigma_{ij}}\int_{-\infty}^{a}\cdots \int_{-\infty}^{a}\exp\left(-\frac{1}{2}\*y^T\*\Sigma^{-1}\*y\right)d\*y \\
&+|\*\Sigma|^{-1/2}(2\pi)^{-m^*/2}\int_{-\infty}^{a}\cdots \int_{-\infty}^{a}\frac{\delta (-\frac{1}{2}\*y^T\*\Sigma^{-1}\*y)}{\delta \Sigma_{ij}}\exp\left(-\frac{1}{2}\*y^T\*\Sigma^{-1}\*y\right)d\*y. \\
\end{array}
\]
We use Jacobi's formula to simplify the first term with $\delta |\*\Sigma|/\delta \Sigma_{ij} = 2|\*\Sigma| (\*\Sigma^{-1})_{ij}$. Noting that $\frac{\delta \*\Sigma^{-1
}}{\delta \Sigma_{ij}} = \*\Sigma^{-1}\frac{\delta \*\Sigma}{\delta \Sigma_{ij}}\*\Sigma^{-1} $ we have that:
\[
\begin{array}{rl}
\frac{\delta f(\*\Sigma)}{\delta \Sigma_{ij}} =& -\left(\*\Sigma^{-1}\right)_{ij}f(\*\Sigma)+|\*\Sigma|^{-1/2}(2\pi)^{-m^*/2}\int_{-\infty}^{a}\cdots \int_{-\infty}^{a}\frac{\delta \left(-\frac{1}{2}\*y^T\*\Sigma^{-1}\*y\right)}{\delta \Sigma_{ij}}\exp\left(-\frac{1}{2}\*y^T\*\Sigma^{-1}\*y\right)d\*y \\
=&  -\left(\*\Sigma^{-1}\right)_{ij}f(\*\Sigma) + E\left(\frac{1}{2}\*y^T  \*\Sigma^{-1}\frac{\delta \*\Sigma}{\delta \Sigma_{ij}}\*\Sigma^{-1} \*y \right) - \{1-f(\*\Sigma)\} h_{ij}(a)\\
=&  -\left(\*\Sigma^{-1}\right)_{ij}f(\*\Sigma) + \frac{1}{2}\mbox{tr}\left(\*\Sigma^{-1}\frac{\delta \*\Sigma}{\delta \Sigma_{ij}}\*\Sigma^{-1} \*\Sigma\right)-\{1-f(\*\Sigma)\} h_{ij}(a) \\
=&  -\left(\*\Sigma^{-1}\right)_{ij}f(\*\Sigma) + \left(\*\Sigma^{-1}\right)_{ij}-\{1-f(\*\Sigma)\} h_{ij}(a)\\
=&\{1-f(\*\Sigma)\}  \left\{\left(\*\Sigma^{-1}\right)_{ij}-h_{ij}(a)\right\} \\
\end{array}
\]
For large $B$, we have that $\mbox{var}(\hat{\Sigma}_{ij}) = (1-\Sigma_{ij}^2)^2/B$ \cite{hotelling1953new}. We can use the first order expansion to obtain the variance of $f\left(\hat{\*\Sigma}\right)$:
$$\mbox{var}\left\{\sqrt{B}f\left(\hat{\*\Sigma}\right)\right\} =\{1-f(\*\Sigma)\}^2 \sum_{i<j}\left[\left\{\left(\*\Sigma^{-1}\right)_{ij}-h_{ij}(a) \right\}\left(1-\Sigma_{ij}^2\right)\right]^2  + o_B(1)$$
\end{proof}

\subsection{Proof of Corollary 3}

\begin{proof}
First, note that $p<.5$ implies that $a>0$ for any $\*\Sigma$. As $h_{ij}(a)$ is the conditional mean of a quadratic form, there exists $c^{\dagger}$ such that $h_{ij}(a) \leq c^{\dagger} a^2$ for all $i \neq j$ and for all $a >0$. As $c_a$ is quadratic in the $h_{ij}(a)$, there exists $c^{\ddagger}$ such that $c_a \leq c^{\ddagger}a^4$ for $a>0$. Finally, as $p=\mbox{pr}\{\cup_k (Z_k^* > a)\}>1-\Phi(a) \geq \frac{\exp(-a^2/2)}{\sqrt{2\pi}}(\frac{1}{a}-\frac{1}{a^3})$, there exists $c^*$ such that $c^{\ddagger}a^4\leq c^*(\log p)^2$ for all $p<0.5$. Along with Equation \ref{varphat} and $\mbox{var}(\tilde{p})=p(1-p)/B$, this completes the proof.
\end{proof}

\label{lastpage}

\end{document}